\documentclass[aps,prd,preprintnumbers,nofootinbib]{revtex4}
\usepackage[dvipdfmx]{graphicx}
\usepackage{bm,latexsym,amsmath,amsthm,amssymb,amsfonts,color,mathrsfs}

\usepackage[normalem]{ulem}

\begin{document}

\title{\uline{}Quasilocal inequalities for attractive gravity probe surface}

\author{${}^{1,2,3}$Tetsuya Shiromizu, ${}^{1,2}$Keisuke Izumi, ${}^{4,5}$Hirotaka Yoshino and ${}^{6}$Yoshimune Tomikawa}

\affiliation{${}^1$Department of Mathematics, Nagoya University, Nagoya 464-8602, Japan}
\affiliation{${}^2$Kobayashi-Maskawa Institute for the Origin of Particles and the Universe, Nagoya University, Nagoya 464-8602, Japan}
\affiliation{${}^3$Institute for Advanced Research, Nagoya University, Nagoya 464-8602, Japan}
\affiliation{${}^4$Department of Physics, Osaka Metropolitan University, Osaka 558-8585, Japan}
\affiliation{${}^5$Nambu Yoichiro Institute for Theoretical and Experimental Physics (NITEP), Osaka Metropolitan University, Osaka 558-8585, Japan}
\affiliation{${}^6$Division of Science, School of Science and Engineering, Tokyo Denki University, Saitama 350-0394, Japan}
%
%
\begin{abstract}
We discuss the local and quasilocal properties of the loosely trapped surface (LTS) and the attractive gravity probe surface (AGPS), 
which have been proposed to characterize the strength of gravity in both strong and weak gravity regions using the mean curvature.
In terms of local mass defined in a region surrounded by the two AGPSs and of Geroch quasilocal mass, 
we present several inequalities concerning 
their size and area, which are of particular interest.
We also propose the improved concepts of the LTS/AGPS, which we call LTS Plus (LTS+) and AGPS Plus (AGPS$+$),
defined in terms of expansions of outgoing and ingoing null geodesic congruences on those surfaces.  
Then, the similar inequalities are proven in terms of appropriately defined local mass and the Hawking quasilocal mass.
\end{abstract}

\maketitle

%
%
\section{introduction} 

In general relativity, we have the celebrated theory of black holes and spacetime singularities \cite{Hawking1973}. The key concept is 
the event horizon, which is defined in a global manner. As a local object associated with the event horizon, a trapped surface or 
an apparent horizon is also important. 
However, it is known that they always exist inside the black hole if the cosmic censorship conjecture holds. 
Therefore, it is worthwhile to introduce an observable object outside a black hole that can characterize the region of strong gravity. 
The photon sphere in the Schwarzschild spacetime is a typical example, and people have started to discuss its generalization 
\cite{Claudel2001, Shiromizu2017, Yoshino2017, Yoshino2020, Siino2022, Amo2024}. 
One of such examples is the loosely trapped surface (LTS), which 
we proposed in Ref. \cite{Shiromizu2017}. This concept was formulated in terms of Riemannian geometry in the sense that it is defined on a spacelike hypersurface. 
Interestingly, under certain conditions, one can show a Penrose-like 
inequality for the LTS, namely, $A \leq 4\pi (3GM)^2$, where $A$ is the area of the LTS and $M$ is the Arnowitt-Deser-Misner 
(ADM) mass \cite{Shiromizu2017}. Later, the LTS was extended to the so-called attractive gravity probe surface (AGPS), 
which also covers the weak gravity region and satisfies a similar areal inequality \cite{Izumi2021, Izumi2023}. 

In this paper, we discuss the local and quasilocal properties of the LTS/AGPS and improvement of these concepts.
Here, ``local'' means that the effective mass is given as a volume integral of energy density,
while ``quasilocal'' means that the mass-like quantity for some closed surface is evaluated 
with the physical or geometrical quantities on that surface. 
First, we prove a quasilocal inequality for a nondimensional quantity on a AGPS, from which an inequality for the local mass 
in the region surrounded by two AGPSs is derived.
This gives a non-trivial constraint for AGPSs in the context of gravitational collapse of 
matters/gravitational waves, as well as for compact objects, and this result may be regarded as a generalization of the study 
of spherically symmetric cases \cite{Bizon1989}. 

Next, we propose improved concepts of LTS/AGPS, which we call them the {\it LTS Plus} (LTS$+$) and {\it AGPS Plus} (AGPS+).
Since the concepts of the LTS/AGPS are defined in terms of mean curvature of a sequence of 2-surfaces in a spacelike hypersurface $\Sigma$,
we have to admit that these definitions strongly depend on the choice of $\Sigma$.  
In order to improve this point, we use the expansions of outgoing and ingoing null geodesic congruences
emitted from 2-surfaces to define LTS+/AGPS+ instead of the mean curvature. Then, we derive the inequalities for the LTS+/AGPS+
similarly to the case of the LTS/AGPS.

The rest of this paper is organized as follows. In Sec. 2, we derive an upper bound for the local mass of the region surrounded by the two LTSs/AGPSs 
and a lower bound for the size of that region. A quasilocal areal inequality is also presented. 
In Sec. 3, we define {LTS+} and AGPS+, associated with the null expansion, and present similar inequalities. 
In Sec. 4, we give a summary. In the Appendix, we present a different formulation for the variant of the AGPS, 
which we call AGPS Minus (AGPS$-$). 

\section{LTS/AGPS}

Let us consider a spacelike hypersurface $(\Sigma,\,q_{ab},\,K_{ab})$, where $q_{ab}$ and $K_{ab}$ are
the induced metric and the extrinsic curvature of $\Sigma$. 
The AGPS is defined as a compact closed $2$-surface $(S_\alpha,\,h_{ab},\,k_{ab})$ embedded in 
$(\Sigma,\,q_{ab},\,K_{ab})$, 
where $h_{ab}$ and $k_{ab}$ are the induced metric and the extrinsic curvature 
associated with the outward directed unit normal vector $r^a$,
satisfying both $k>0$ and $r^aD_ak \geq \alpha k^2$, where $k$ is the trace of the 
extrinsic curvature , $D_a$ is the covariant 
derivative with respect to the metric $q_{ab}$ and $\alpha$ is a constant greater than $-1/2$ \cite{Izumi2021}. 
In the special case $\alpha=0$, the surface is called the LTS \cite{Shiromizu2017}. 

As derived in \cite{Shiromizu2017}, the relation
\begin{eqnarray}
r^aD_ak=-\frac{1}{2}k^2-\frac{1}{2}k_{ab}k^{ab}+\frac{1}{2}{}^{(2)}R-\frac{1}{2}{}^{(3)}R-\varphi^{-1}{\cal D}^2\varphi
\label{rDk}
\end{eqnarray}
holds, where ${\cal D}$ is the covariant derivative with respect to $h_{ab}$, ${}^{(2)}R$ is the Ricci scalar of $S_\alpha$, ${}^{(3)}R$ is the Ricci scalar of $\Sigma$, and $\varphi$ 
is the lapse function associated with $r^a$. Using Eq.~\eqref{rDk}, one of the AGPS conditions $r^aD_ak \geq \alpha k^2$ implies
\begin{eqnarray}
-\Bigl( \alpha+\frac{3}{4} \Bigr)k^2-\frac{1}{2}\tilde k_{ab} \tilde k^{ab}+\frac{1}{2}{}^{(2)}R-\frac{1}{2}{}^{(3)}R-\varphi^{-1}{\cal D}^2\varphi
 \geq 0,
\end{eqnarray}
where $\tilde k_{ab}$ is the traceless part of $k_{ab}$. 
If ${}^{(3)}R \geq 0$ is satisfied everywhere on $S_\alpha$,  integration over $S_\alpha$ yields
\begin{eqnarray}
0<\int_{S_\alpha} \Biggl[ \Bigl( \alpha+\frac{3}{4} \Bigr)k^2+\frac{1}{2}\tilde k_{ab}\tilde k^{ab}+\frac{1}{2}{}^{(3)}R
+\varphi^{-2}({\cal D}\varphi)^2 \Biggr] dA \leq \frac{1}{2}\int_{S_\alpha} {}^{(2)}RdA = 4\pi.  \label{original}
\end{eqnarray}
In this inequality, we used the fact that, through the Gauss-Bonnet theorem, the positivity of $\int_{S_\alpha} {}^{(2)}RdA$ implies that the topology of $S_\alpha$ is that of 2-sphere, and consequently, the integral equals to  $4\pi$. 
This inequality leads to several consequences. First, we have 
\begin{eqnarray}
\int_{S_\alpha} \Biggl( \frac{1}{2}\tilde k_{ab}\tilde k^{ab}+\frac{1}{2}{}^{(3)}R 
 \Biggr) dA \leq 4\pi. \label{ub}
\end{eqnarray}
Second, we obtain an upper bound for the Willmore energy $\mathcal{W}(S_\alpha)$\footnote{
Note that 
although $4\pi$ is known as a lower bound for the Willmore energy, $\mathcal{W} \geq 4\pi$, for a surface immersible into 3-Euclidean space $\bm{E}^3$ \cite{Willmore1965}, 
this bound is not applicable to the AGPS in a general curved spacelike hypersurface. 
}:
\begin{eqnarray}
\mathcal{W}(S_\alpha) \leq \pi\Bigl( \alpha+\frac{3}{4} \Bigr)^{-1}, \label{wilmore}
\end{eqnarray}
where $\mathcal{W}(S_\alpha)$ is defined by 
\begin{equation}
\mathcal{W}(S_\alpha)\, := \, \frac{1}{4} \int_{S_\alpha} k^2 dA.
\end{equation}
Third, we can show the positivity of the Geroch mass $m_G(S_\alpha)$ as (see also Eq.~(8) in Ref. \cite{Shiromizu2024}) 
\begin{eqnarray}
m_G(S_\alpha) & := & \frac{1}{16 \pi G} \Bigl( \frac{A}{4 \pi}\Bigr)^{1/2} \int_{S_\alpha} \Biggl({}^{(2)}R-\frac{1}{2}k^2 \Biggr)dA \nonumber \\
& = &  \frac{1}{8 \pi G} \Bigl( \frac{A}{4 \pi}\Bigr)^{1/2} \int_{S_\alpha} \Biggl[r^aD_ak +\frac{1}{2}k^2
+\frac{1}{2}{}^{(3)}R+\frac{1}{2}\tilde k_{ab}\tilde k^{ab}+\varphi^{-2}({\cal D} \varphi)^2   \Biggr]dA \nonumber \\
& \geq &  \frac{1}{8 \pi G} \Bigl( \frac{A}{4 \pi}\Bigr)^{1/2} \int_{S_\alpha} \Bigg[ \Bigl( \alpha+\frac{1}{2} \Bigr)k^2
+\frac{1}{2}{}^{(3)}R+\frac{1}{2}\tilde k_{ab}\tilde k^{ab}+\varphi^{-2}({\cal D} \varphi)^2   \Biggr]dA \nonumber \\
& > &  0. \label{geroch}
\end{eqnarray}

Suppose $\Sigma$ to be a maximal hypersurfaces embedded in a spacetime $(M,\,g_{ab})$.
Using the Einstein equation and the Gauss equation, the scalar curvature ${}^{(3)}R$ can be expressed as
\begin{eqnarray}
{}^{(3)}R & = & 16\pi G \rho +K_{ab}K^{ab} \nonumber \\
          & = & 16\pi G \rho +2v_av^a+\kappa_{ab} \kappa^{ab}+K_{(r)}^2 \nonumber \\
          &  \geq &  16\pi G \rho +2v_av^a+\tilde \kappa_{ab} \tilde \kappa^{ab},
\end{eqnarray}
where $\rho:=T_{ab}n^an^b$ is the energy density of matters, $n^a$ is the timelike unit normal vector to $\Sigma$, 
$K_{ab}$ is the extrinsic curvature of $\Sigma$ in $M$, 
and other quantities are defined as $K_{(r)}:=K_{ab}r^ar^b$ and $\kappa_{ab}:={h_a}^c{h_b}^dK_{cd}$, 
$v_a:={h_a}^bK_{bc}r^c$. 
$\tilde \kappa_{ab}$ is the traceless part of $\kappa_{ab}$. 
The vector field $v^a$ is naturally related to the angular momentum.
Then, under the assumption $\rho \geq 0$, Eq.~(\ref{ub}) holds and leads to the inequality
\begin{eqnarray}
\int_{S_\alpha} \Biggl(8\pi G \rho+ \frac{1}{2}\tilde k_{ab}\tilde k^{ab}+\frac{1}{2}\tilde \kappa_{ab}\tilde  \kappa^{ab}+v_av^a 
\Biggr) dA \leq 4\pi. \label{ub-max}
\end{eqnarray}
One may define the energy density for gravitational wave as 
$8\pi G \rho_{\rm gw}:=(1/2)(\tilde k_{ab}\tilde k^{ab}+\tilde \kappa_{ab}\tilde \kappa^{ab})$ and the effective pressure 
from the angular momentum as $8 \pi G p_{\rm ang}:=v_av^a$. Substituting these definitions into Eq.~(\ref{ub-max}), we have a quasilocal inequality
for a nondimensional quantity as
\begin{eqnarray}
2 G \int_{S_\alpha} (\rho+\rho_{\rm gw}+p_{\rm ang} ) dA \leq 1. \label{ub-max2}
\end{eqnarray}
This inequality may be regarded as a constraint on the effective energy density on the LTS/AGPS. 
Note, for example, that it becomes 
trivial if $S_\alpha$ is an $r$-constant surface in the time symmetric hypersurface of the Schwarzschild spacetime,
where $\rho=\rho_{\rm gw}=p_{\rm ang}=0$.
If the effective energy density on some surface is too large to violate the condition of Eq.~\eqref{ub-max2}, that surface cannot be an AGPS because
it contradicts the condition $\alpha>-1/2$.

Let us assume that there exists a region $\Omega$, foliated by LTSs/AGPSs, such that $\varphi=1$ in $\Omega$.  
Note that this choice of the lapse function $\varphi$ is not restrictive, because
along the flow of $S^2$ surface defined by $\varphi=1$, the quantity $r^aD_ak$ becomes a continuous function of position
at least in some range of the radial coordinate, and it is easy to find the value of $\alpha$ that satisfies the condition for the AGPS for each $S^2$ surface.
Then, integration of Eq.~(\ref{ub-max2}) over $\Omega$  gives
\begin{eqnarray}
2G \Delta M_{\rm eff} \leq  \Delta L, \label{IHC}
\end{eqnarray}
where the effective mass $\Delta M_{\rm eff}$ of the region $\Omega$ is defined by
\begin{eqnarray}
\Delta M_{\rm eff}:= \int_\Omega (\rho+\rho_{\rm gw}+p_{\rm ang})d\Sigma
\end{eqnarray}
and $\Delta L$ is the geodesic distance between the inner and outer boundary surfaces of $\Omega$.
The physical meaning of this inequality is clear: the width of $\Omega$ is larger than the Schwarzschild 
radius determined by the effective mass $\Delta M_{\rm eff}$ contained in $\Omega$. 
See Ref.~\cite{Bizon1989} for a similar argument in the case of spherically symmetric trapped surface.

It is also worth considering the second inequality of Eq.~(\ref{wilmore}). 
Similarly to the discussion above, we focus on the region $\Omega$. 
By the Cauchy-Schwarz inequality, we have 
\begin{eqnarray}
\left(\int_{S_\alpha} k\, dA \right)^2 \le \left(\int_{S_\alpha} k^2 dA  \right)\left(\int_{S_\alpha}  dA  \right) \le 
 4\pi  \left(\alpha+\frac34\right)^{-1} A.  
\label{CS}
\end{eqnarray}
Since  $r^aD_aA =\int_{S_\alpha}kdA$ holds in the present coordinate system\footnote{
When evaluating one of the AGPS conditions, $ r^a D_a k \ge \alpha k^2$, we have already introduced a foliation near $ S_\alpha $. Here, $ A $ denotes the area of each leaf of the foliation, and $r^a D_a A$  represents its derivative.
}, Eq.~(\ref{CS}) implies 
\begin{eqnarray}
r^a D_a \left(A^{1/2}\right) \leq \pi^{1/2}\left(\alpha+\frac34\right)^{-\frac12}. \label{CS2}
\end{eqnarray} 
For large values of $\alpha$, the right-hand side of Eq.~(\ref{CS2}) becomes smaller, indicating a stronger suppression of area growth along the surface normal.  
This behavior is characteristic of regions near a gravitational source, where displacements 
along the surface normal lead to only small changes in area. In the Schwarzschild case, 
this reflects the stretching of the radial direction near the black hole, where a unit 
radial displacement alters the area much less than the case of flat spacetime. The derived inequality
 can thus be regarded as a general expression of this feature in arbitrary spacetimes, 
indicating that the definition of an AGPS effectively captures the suppression of area growth 
in the normal direction as a manifestation of gravitational strength. 
The integration of Eq.~(\ref{CS2}) over $\Omega$ yields
\begin{eqnarray}
A^{1/2}_o-A^{1/2}_i < 2 \pi^{1/2} \Delta L, \label{ineq}
\end{eqnarray} 
where $A_o$ and $A_i$ denote the areas of the outer and inner boundaries of $\Omega$, respectively. Here we have used 
$\alpha+3/4 > 1/4$, which follows from the restriction on $\alpha$ in the definiton of the AGPS. 

For the convenience in the next section, here we recall the quasilocal inequality on the AGPSs in terms of the Geroch mass derived in
Refs.~\cite{Shiromizu2017,Izumi2021, Izumi2023}.
Combining the second inequality of Eq.~(\ref{original}) with the definition of the Geroch mass, we have 
\begin{eqnarray}
m_G (S_\alpha) \geq \frac{1}{G}\frac{2\alpha+1}{4\alpha+3} \Bigl(\frac{A}{4\pi} \Bigr)^{1/2}.
\end{eqnarray} 
Then, we obtain the quasilocal areal inequality 
\begin{eqnarray}
A \leq 4 \pi \Biggl( \frac{4\alpha+3}{2\alpha+1}G m_{G}(S_\alpha) \Biggr)^2. 
\end{eqnarray} 
If one can show the monotonicity of the Geroch mass \cite{Geroch1973}, one can prove the Penrose-like inequality for the LTS/AGPS
in terms of the ADM mass \cite{Shiromizu2017, Izumi2021, Izumi2023}.
In the next section, we present a similar inequality for the AGPS$+$ in terms of the Hawking quasilocal mass instead of the
Geroch mass.


%


\section{LTS/AGPS plus} 

LTSs/AGPSs are originally defined within a spacelike hypersurface, and thus their definitions are expressed in terms of geometric quantities intrinsic to such a hypersurface.
Although it has turned out that LTSs/AGPSs properly characterize the strength of gravity on maximally sliced spacelike hypersurfaces, 
for physical applications, it is desirable to formulate these definitions in a manner less dependent on the choice of the spacelike hypersurfaces.
In this section, we propose one possible extension of the LTS/AGPS definition using the expansions $\theta_\pm$ of outgoing and ingoing null geodesic congruences from the surface, and examine whether the characteristic features discussed in Sec.~2 are preserved.

Let us consider a compact $2$-surface ($S$,\,$h_{ab}$,\,$k_{ab}$) embedded in a $3$-dimensional 
spacelike hypersurface ($\Sigma$,\,$q_{ab}$,\,$K_{ab}$), which is embedded in a $4$-dimensional 
spacetime ($M$,\,$g_{ab}$). Denoting the future-directed unit normal vector to $\Sigma$ in $M$ by $n^a$
and the outward-directed spacelike unit normal vector to $S$ in $\Sigma$ by $r^a$, 
the metric $g_{ab}$ is decomposed as 
\begin{equation}
g_{ab}=h_{ab}+r_ar_b-n_an_b=q_{ab}-n_an_b. 
\end{equation}

We now define the AGPS Plus (AGPS+) as a compact $2$-surface $S$ in a spacelike hypersurface $\Sigma$ 
satisfying both $\theta_+>0$ and $r^a \nabla_a \theta_+ \ge -\alpha \theta_+ \theta_-  $, where $\theta_+=h^{ab} \nabla_a k_b$ is the null expansion 
associated with the outward directed null vector $k^a:=n^a+r^a$ to $S$, 
$\theta_-=h^{ab} \nabla_a l_b$ is the null expansion 
associated with the inward directed null vector $l^a:=n^a-r^a$ to $S$, 
and $\alpha$ is a constant satisfying $\alpha >-1/2$. 
In the special case $\alpha=0$, the surface is referred to as an LTS Plus (LTS+).
On the time-symmetric initial data, this definition reduces to that for the AGPS. 
In Appendix of Ref.~\cite{Shiromizu:2022}, the following equation has been derived:
\begin{eqnarray}
r^a \nabla_a \theta_+ =-\frac{1}{2}\theta_{+ab}\theta_+^{ab}-\frac{1}{2}\theta_+^2+\theta_+K+\frac{1}{2}{}^{(2)}R+{\cal D}_aV^a
-V_aV^a -G_{ab}k^an^b, \label{rDtheta}
\label{Eq-derivative-of-thetaplus}
\end{eqnarray}
where $\theta_{+ab}={h_a}^c{h_b}^d \nabla_c k_d$, ${}^{(2)}R$ is the Ricci scalar of $S$, $K$ is the trace 
part of the extrinsic curvature $K_{ab}$ of $\Sigma$ and $G_{ab}$ is the 
Einstein tensor. 
$V_a$ is defined by $V_a=v_a-{\cal D}_a \ln \varphi$, where $v_a:={h_a}^b K_{bc}r^c $ is related to the 
angular momentum and $\varphi$ is the lapse function associated with $r^a$. 

When $\Sigma$ is a maximal spacelike hypersurface, {\it i.e.} $K=0$, Eq.~(\ref{rDtheta}) with the LTS+ condition yields 
\begin{eqnarray}
\frac{1}{2}\theta_{+ab}\theta_+^{ab}+\frac{1}{2}\theta_+^2-{\cal D}_aV^a
+V_aV^a +G_{ab}k^an^b \leq \frac{1}{2}{}^{(2)}R. 
\label{LTS+ineq}
\end{eqnarray}
Assuming $G_{ab}k^an^b\ge0$, which follows from the dominant energy condition, the integration of the inequality of Eq.~\eqref{LTS+ineq} over $S$ implies 
\begin{eqnarray}
\int_S \Bigl( \frac{1}{2}\tilde \theta_{+ab} \tilde \theta_+^{ab}+V_aV^a +G_{ab}k^an^b \Bigr)dA  < \frac{1}{2}\int_S {}^{(2)}RdA =4\pi, 
\end{eqnarray}
where we have used the same reasoning as that in Eq.~\eqref{original}.
Here, $\tilde \theta_{+ab}$ is the traceless part of $\theta_{+ab}$. One may define the energy deinsity for 
gravitational waves as 
\begin{eqnarray}
\rho_{\rm +gw}:=\frac{1}{16\pi G}\tilde \theta_{+ab} \tilde \theta_+^{ab}
\end{eqnarray}
and the effective pressure associated with $V_a$ as 
\begin{eqnarray}
p_{\rm +ang}:=\frac{1}{8\pi G} V_aV^a.
\end{eqnarray}
Using the Einstein equation, we have
\begin{eqnarray}
8 \pi G \int_S \Bigl( \rho_++\rho_{\rm +gw} +p_{\rm +ang} \Bigr)dA  \leq \frac{1}{2}\int_S {}^{(2)}RdA =4\pi, \label{lts-plus1}
\end{eqnarray}
where $8\pi G \rho_+:=G_{ab}k^an^b$. 

Assume that, on the maximal spacelike hypersurface, there exists a region $\Omega$ foliated by LTS+ surfaces
where the lapse of the radial coordinate is $\varphi=1$ in $\Omega$. 
Note that this condition is not very restrictive, because if one can find one LTS+,
there must exist a foliation of $S^2$ surfaces determined by $\varphi=1$ at least locally, and 
the quantities $r^a\nabla_a\theta_+$ and $\theta_+$ must become continuous functions.  
Then, integrating over $\Omega$, we obtain  
\begin{eqnarray}
2G \Delta {\cal M}_{\rm +eff} \leq \Delta L, \label{IHC2}
\end{eqnarray}
where 
\begin{eqnarray}
\Delta {\cal M}_{\rm +eff} := \int_\Omega (\rho_++\rho_{\rm +gw} +p_{\rm +ang})d \Sigma
\end{eqnarray}
and $\Delta L$ is the geodesic distance between the inner and outer boundaries of $\Omega$. 
Note that although we considered a sequence of the LTS$+$'s in the above inequalities, the same inequalities
hold for a sequence of the AGPS$+$'s as long as $\theta_+ > (4/3)\alpha\theta_-$ is satisfied.

We are able to show the Penrose-like inequality for the AGPS+ in terms of the 
Hawking quasilocal mass \cite{Hawking1968},
\begin{equation}
  m_{\rm H}(S) \, = \, \frac{1}{16\pi G} \Bigl( \frac{A}{4\pi}\Bigr)^{1/2} \int_{S} \Biggl({}^{(2)}R+\frac{1}{2}\theta_+ \theta_- \Biggr)dA.
\end{equation}
Rewriting 
Eq.~(\ref{rDtheta}) as 
\begin{eqnarray}
r^a \nabla_a \theta_+ =-\frac{1}{2}\tilde \theta_{+ab} \tilde \theta_+^{ab}+\frac{3}{4}\theta_+ \theta_--\frac{1}{2}(\kappa-2K_{(r)}) \theta_+
+\frac{1}{2}{}^{(2)}R+{\cal D}_aV^a-V_aV^a -G_{ab}k^an^b, \label{rDtheta2}
\end{eqnarray}
where $\kappa$ is the trace of $\kappa_{ab}$ and
we used the fact that $\theta_\pm =\kappa\pm k=K-K_{(r)} \pm k$, imposing the additional condition\footnote{On a maximal slice 
$K=\kappa+K_{(r)}=0$,  the inequality of Eq.~\eqref{kappa-2K} requires $\kappa \geq 0$, that is, the area of $S_\alpha$ momentarily increases toward the future.} 
\begin{eqnarray}
\kappa-2K_{(r)} \geq 0
\label{kappa-2K}
\end{eqnarray} 
for the AGPS+, and assuming $\rho_+\ge 0$, we find that 
\begin{eqnarray}
\int_{S} \Biggl[\Bigl(\frac{3}{4}+\alpha \Bigr)\theta_+\theta_-+\frac{1}{2}{}^{(2)}R   \Biggr]dA \geq 0. \label{agpsplus}
\end{eqnarray} 
Assuming that $ \int_{S} \theta_+ \theta_- \, dA \leq 0 $, a short manipulation then shows that the Hawking mass is bounded from below:
\begin{eqnarray}
m_{\rm H}(S)
& \geq & \frac{1}{G}\frac{2\alpha+1}{4\alpha+3} \Bigl(\frac{A}{4\pi} \Bigr)^{1/2}
\end{eqnarray} 
which leads to the areal inequality for the AGPS+ as
\begin{eqnarray}
A \leq 4 \pi \Biggl( \frac{4\alpha+3}{2\alpha+1}G m_{\rm H}(S) \Biggr)^2. \label{PIplus}
\end{eqnarray} 
For the LTS+ ($\alpha=0$), this inequality becomes
\begin{eqnarray}
A \leq 4 \pi \Biggl( 3G m_{\rm H}(S) \Biggr)^2. 
\end{eqnarray} 
The monotonicity of the Hawking mass along several kind of flows has been discussed,
e.g. in Refs.~\cite{Frauendiener:2001,Bray:2006,Hirsch:2022}. 
In the case that such a flow exists, the Hawking mass in the quasilocal Penrose-like inequality can 
be replaced by the ADM mass.


\section{summary}

In this paper, we have derived the inequality of Eq.~(\ref{original}) for the loosely trapped surface (LTS) and attractive gravity probe surface (AGPS) in spacelike hypersurface $\Sigma$ with non-negative Ricci scalar, 
which leads to several comprehensive consequences. 
In particular, 
we focused on the consequences of the inequality of Eq.~(\ref{ub}). 
Using the Einstein equation, we obtained the inequality of Eq.~(\ref{ub-max2}), which constrains a nondimensional quantity on the LTS/AGPS. 
Furthermore, the width of the region between two LTSs/AGPSs, equipped with the normal coordinate, 
satisfies the inequality of Eq.~(\ref{IHC}). This is quite natural because the size outside a black hole must be larger than the Schwarzschild radius 
determined by the effective mass $\Delta {\cal M}_{\rm eff}$. 
We have also proved the lower bound for the width of a region $\Omega$ surrounded by two AGPSs 
in terms of the areas of those AGPSs as presented in  Eq.~\eqref{ineq}.

As a trial study, we have introduced the AGPS+ in terms of spacetime geometry. 
On a maximal spacelike hypersurface, this led to the inequality of Eq.~(\ref{lts-plus1}) for a nondimensional quantity on AGPS$+$. 
Similarly to the inequality of Eq.~(\ref{IHC}),  we have also derived the inequality of Eq.~(\ref{IHC2})
that gives a constraint on the width of a region surrounded by two AGPS$+$'s and the local mass in that region. 
We have also obtained the Penrose-like inequality for the AGPS$+$ in terms of the Hawking quasilocal mass, just as for the AGPS.

%
%
\begin{acknowledgments}

T. S., K. I. and H. Y. are supported by Grant-Aid for Scientific Research from Ministry of Education, Science, 
Sports and Culture of Japan JSPS(No. JP21H05189). 
T. S.  and  K. I. are supported by Grant-Aid for Scientific Research from Ministry of Education, Science, 
Sports and Culture of Japan JSPS(JP21H05182). 
T. S. is also supported by JSPS Grants-in-Aid for 
Scientific Research (C) (JP21K03551), Fund for the Promotion of Joint International Research (JP23KK0048)
and Grant-in-Aid for Scientific Research (A) (JP24H00183). 
K. I. is also supported by JSPS Grant-in-Aid for Scientific Research (C) (JP24K07046). 
H.Y is in part supported by JSPS KAKENHI Grant Numbers JP22H01220, and is partly supported by Osaka Central 
Advanced Mathematical Institute (MEXT Joint Usage/Research Center on Mathematics and Theoretical Physics JPMXP0619217849). 
\end{acknowledgments}

\begin{appendix}

\section{Another formulation for AGPS: AGPS Minus}

Here we discuss another formulation for the extension of the AGPS,  the AGPS Minus (AGPS$-$) 
in terms of the ingoing null expansion.  
It is defined as a compact $2$-surface $S$ in a spacelike hypersurface $\Sigma$ 
satisfying $\theta_-<0$ and $r^a \nabla_a \theta_- < \alpha \theta_+ \theta_- $, where $\alpha$ is a constant satisfying $\alpha >-1/2$. Note that 
this definition reduces to the original AGPS on time-symmetric initial data. 
In the similar manner to Eq.~\eqref{Eq-derivative-of-thetaplus},
we can derive the following key equation: 
\begin{eqnarray}
r^a \nabla_a \theta_- & = & \frac{1}{2} \theta_{-ab} \theta_-^{ab}+\frac{1}{2} \theta_-^2-K \theta_-+{\cal D}_a W^a+W_aW^a-\frac{1}{2} {}^{(2)}R+G_{ab}n^a \ell^b \\
& = & \frac{1}{2}\tilde \theta_{-ab} \tilde \theta_-^{ab}-\frac{3}{4} \theta_+ \theta_-+\frac{1}{2} ( \kappa-2K_{(r)} ) \theta_-
+{\cal D}_a W^a+W_aW^a-\frac{1}{2} {}^{(2)}R+G_{ab}n^a l^b, 
\end{eqnarray} 
where $\theta_{-ab}:={h_a}^c{h_b}^d \nabla_c l_d$, $W_a:=v_a+{\cal D}_a \ln \varphi$, and $\tilde\theta_{-ab}$ is the traceless part of $\theta_{-ab}$. 

For the LTS$-$ on a maximal slice, we obtain  
\begin{eqnarray}
2G \int_{S_\alpha} (\rho_{-{\rm gw}}+p_{-{\rm ang}}+\rho_-) dA <1,
\end{eqnarray} 
where 
\begin{eqnarray}
8\pi G \rho_{-{\rm gw}}:=\frac{1}{2}\tilde \theta_{-ab} \tilde \theta_-^{ab},~~8\pi G p_{-{\rm ang}}:=W_aW^a,~~8 \pi G \rho_-:=G_{ab}n^a \ell^b.
\end{eqnarray} 
In the same setup as that of the main text except that LTS+ is replaced by LTS$-$, the integration of the above inequality in the region $\Omega$ yields
\begin{eqnarray}
2G \Delta {\cal M}_{-{\rm eff}}< \Delta L,
\end{eqnarray} 
with 
\begin{eqnarray}
\Delta {\cal M}_{-{\rm eff}}:= \int_{\Omega} (\rho_{-{\rm gw}}+p_{-{\rm ang}}+\rho_-) d\Sigma,
\end{eqnarray} 
where $\Delta L$ is the geodesic distance between the inner and outer boundaries of $\Omega$. 

On a slice satisfying the additional condition $2K_{(r)}-\kappa \geq 0$ \footnote{On a maximal slice, this requires $\kappa \leq 0$, 
which means that the area of $S_\alpha$ momentarily decreases toward the future.}, we have Eq.~(\ref{agpsplus}), and hence  
the quasilocal areal inequality of Eq.~(\ref{PIplus}) holds. 

\end{appendix}


\end{document}